\documentstyle[aps,twocolumn,floats]{revtex}
\draft
\begin{document}
\newcommand{\bec}{\begin{center}}
\newcommand{\ec}{\end{center}}
\newcommand{\be}{\begin{equation}}
\newcommand{\ee}{\end{equation}}
\newcommand{\beqn}{\begin{eqnarray}}
\newcommand{\eeqn}{\end{eqnarray}}
\newcommand{\bet}{\begin{table}}
\newcommand{\ent}{\end{table}}
\newcommand{\bib}{\bibitem}

\title{Orbital Dependent Exchange-Only Methods for Periodic Systems}

\author{P. S\"ule$^1$, S. Kurth$^2$ and V. E. Van Doren$^1$} 
\address{
  $1$ Department Natuurkund,
University of Antwerp (RUCA)\protect\\, Groenenborgerlaan 171.,\\ B-2020 Antwerpen, Belgium \protect\\
 E-mail: sule@ruca.ua.ac.be\protect\\
 $2$ Dept. of Physics and Quantum Theory Group\protect\\
 Tulane University, New Orleans, LA 70118 USA}

\maketitle

\begin{abstract}

\wideabs{

 Various orbital-dependent exchange-only potentials are studied which exhibit
correct long-range asymptotic behaviour.
We present the first application of these potentials for polymers and by one of these potentials for molecules.
Kohn-Sham type calculations have been carried out for 
polyethylene in order to make valuable comparison of these potentials with each
other as well as with Hartree-Fock and exchange-only LDA ($X\alpha$) methods. 
The difference between total energies and highest occupied orbital energies obtained with Hartree-Fock methods and with localized exchange-potentials
is larger for this polymer than for atoms or molecules.
Various properties of the band structure are also calculated. The band gap 
strongly depends on the basis set. The larger basis set makes the Kohn-Sham eigenvalue gap too low
at about $4.4$ eV
while the minimal basis set results in value close to the experimental
gap ($\sim 8.8$ eV).
For the total energy and the exchange energy,
the various orbital-dependent exchange-only and Hartree-Fock results differ only slightly,
but for the highest occupied orbital energy the difference is more pronounced.
The Kohn-Sham band gap obtained with the optimized effective potetial method is
corrected 
with the exchange contribution to the derivative discontinuity of the exchange-correlation
potential. The corrected band gap obtained with the Slater's exchange potential
is $9.7$ eV close to the experiment.
}

\end{abstract}

\baselineskip 4.0mm

{\bf 1. Introduction}\\

\vspace{5mm}

   Density Functional Theory (DFT) is well established in theoretical physics
and quantum chemistry \cite{DFT}.
The application of the theory to investigate the properties of materials
has been made in the framework of the so-called second generation density
functionals in which the kinetic energy is expressed in terms of orbitals
but the exchange and correlation energy $E_{xc}$ in terms of the electron
density. This leads to the one-electron Kohn-Sham (KS) equations. In many
cases $E_{xc}$ is treated in the local density approximation (LDA).
However, for atoms and molecules gradient correction to the electron
density, such as the generalized gradient approximation (GGA) is necessary
in order to improve the results. 

  Density functionals of the third generation treat both the kinetic and 
exchange energy exactly in terms of single-particle orbitals and only the
correlation energy needs to be approximated, either in terms of the density
or in terms of the orbitals.
The corresponding one-particle equations are known as the
optimized
effective potential method (OEP) derived for the first time quite some time ago by Sharp and
Horton and later by Talman and Shadwick \cite{Sharp,TS}.
These OEP integro-differential equations yield orbitals which are implicit
functionals of the density $\rho$ since the orbitals come from a local
potential. 
Therefore, one can use the following notation for such a class of potentials:
$v_x[\{u_i([\rho],{\bf r})\}]$.
The Hohenberg-Kohn theorem \cite{HK}, applied to noninteracting systems,
ensures that the ground-state determinant, hence, all the occupied
orbitals are unique functionals of the density.

  However, the OEP method is still rather complicated. Krieger, Li and
Iafrate (KLI) transformed and approximated these equations into a manageable
form and applied them to atoms \cite{KLI,KLIatom} where they yield results nearly
identical to OEP \cite{KLI}.
Later the KLI approach was also used for molecules \cite{KLImol} and for
certain semiconductors \cite{Kotani,Bylander}, but not for polymer chains.
Other calculations with a Kohn-Sham potential which is believed to be close
to the exact KS potential resulted in much narrower gap than the experiment \cite{Godby}.

 The KLI approximation to OEP provides a self-interaction free exchange potential and proper $-\frac{1}{r}$ asymptotic behaviour as $r \rightarrow \infty$ \cite{KLI}. 
 Other approximate potentials $v_x[\{u_i\}]$ are reported which exhibit proper asymptotics as well \cite{Gritsenko}. 
The Koopmans theorem \cite{KLI} is satisfied by the highest occupied orbital in OEP and KLI,
while it is violated by other density functionals such as LDA and GGA.
For instance, while the popular GGA exchange-correlation energy functional \cite{PW91}
provides large improvements in the relative  energies of various systems, the exchange-correlation
potentials and the energy eigenvalues of the highest occupied state, are both suffering
from significant error \cite{KLI}.
Also others found insignificant improvement to semiconductor energy gap by
GGA \cite{Ortiz}.
 In the last decades quite a number of publications appeared on
polymer electronic structure using mainly Hartree-Fock {\em ab initio}
methods \cite{Ladik}. Results are reported only the most recently,
however, using DFT methods like LDA
\cite{DFT,LDApolymer}.

 In this paper we will present results obtained by various orbital dependent local exchange potentials for polymers. We give the valuable comparison of different $v_x$ potentials
which allows one to draw definite conclusions about the future applicability of OEP based methods for
periodic systems. 
The work presented here describes the
 calculated band gaps, band widths, band structures and total energy of polyethylene
within the Kohn-Sham density functional scheme using, however, various orbital dependent
exchange potentials as the Slater's potential and the reasonably
simplified version of the KLI exchange-only potential. For comparison we also
give results obtained by Hartree-Fock (HF) and simple periodic exchange-only  
LDA (X$\alpha$).

\vspace{5mm}

{\bf 2. Basic formalism}\\ 

 We start from the following Kohn-Sham effective single-particle equation
of ordinary density functional theory (for sake of simplicity we dropped the spin index and atomic units are used throughout the paper):
\be
[-\frac{1}{2} \nabla^2+ v_s({\bf r})] u_i({\bf r})=
 \epsilon_i u_i({\bf r})
\label{kseq}
\ee
where $\{u_i, i=1,2,...,N\}$ are eigenfunctions of the single particle effective
KS potential $v_s({\bf r})$ \cite{DFT}.
$i$ is a collective index for all single particle quantum numbers. 
\be
v_s({\bf r})=v_{ext}({\bf r})+v_H({\bf r})+v_{xc}({\bf r})
\label{kspot}
\ee
As usual, $v_{ext}({\bf r})$ denotes the external and $v_H({\bf r})$ the Hartree potential. 
$v_{xc}({\bf r})$ is a {\em local} exchange-correlation potential
which is formally defined as the functional derivative of the exchange-correlation
energy
\be
v_{xc}({\bf r},[\rho])=\frac{\delta E_{xc}[\rho]}{\delta \rho({\bf r})}.
\label{funcder}
\ee
Due to the Hohenberg-Kohn theorem, there exists a one-to-one mapping 
between the single particle KS potentials $v_s({\bf r})$ and densities
$\rho({\bf r})$ which guarantees that the functional derivative
given by Eq.~(\ref{funcder}) is defined \cite{DFT}.

The exact $E_{xc}[\rho]$ can be written as 
\be
E_{xc}[\rho]=\frac{1}{2} \int d{\bf r}' d{\bf r} \int_0^1 d\lambda
\frac{\rho({\bf r}) \rho({\bf r}') [g^{\lambda}([\rho];{\bf r},{\bf r'})-1]}
 {\vert {\bf r}-{\bf r}' \vert},
\ee
where $g^{\lambda}([\rho];{\bf r},{\bf r'})$ is the pair correlation
function of the ficticious system 
with interaction strength parameter $0 \le \lambda \le 1$ and a ground
state density which is independent of $\lambda$.
When employing Eq.~(\ref{funcder}) on $E_{xc}[\rho]$,
\beqn
v_{xc}({\bf r},[\rho])=\int_0^1 d\lambda \biggm\{ \int d{\bf r}' \frac{\rho({\bf r}') [g^{\lambda}([\rho];{\bf r},{\bf r}')-1]}
{\vert {\bf r}-{\bf r}^{'} \vert } && \\ \nonumber 
 +\frac{1}{2} \int d{\bf r}' d{\bf r}^{''} \frac{\rho({\bf r}') \rho({\bf r}^{''})}
{\vert {\bf r}'-{\bf r}^{''} \vert} \frac{\delta g^{\lambda}([\rho];{\bf r}',{\bf r}^{''}])}{\delta \rho({\bf r})} \biggm\}.
\label{vxder}
\eeqn
The first term of the potential is the so-called {\em potential} part of the
exchange-correlation energy \cite{Sule,Gritsenko} which is identical with Slater's potential in the exchange-only case, i.e. for $g^{\lambda}$ approximated by $g^{\lambda=0}$ and which is exactly known
\cite{Slater}. 
 To get the second, {\em response}-like term on the right-hand side of Eq. (5) one faces
the technical difficulty that - even for the exchange-only case - the functional derivative can not be determined directly.
One way to get $v_{xc}({\bf r})$ for a given, orbital-dependent approximation to $E_{xc}$ is provided by
the optimized effective potential method (OEP)
 given by Talman and Shadwick \cite{TS}.
The starting point of the OEP method is the total energy functional
\beqn
\enspace
E_{tot}^{OEP}[\rho] = \sum_{i=1}^{occ} \int d{\bf r} u_i^*({\bf r}) \biggm(
-\frac{1}{2} \nabla^2 \biggm) u_i({\bf r}) \nonumber \\
+\int d{\bf r} \rho({\bf r}) v_{ext}({\bf r}) 
+ \frac{1}{2} \int d{\bf r} d{\bf r}' \frac{\rho({\bf r}) \rho({\bf r}')}
{\vert \bf{r}-\bf{r}' \vert} 
+ E_{xc}^{OEP}[\{u_i\}],
\eeqn
In contrast to ordinary DFT, the exchange-correlation energy is an explicit
functional of orbitals and therefore only an implicit functional of the 
density via Eq.~(\ref{kspot}) \cite{KLI,KLImol}.
The local single-particle potential appearing in Eqs.~(\ref{kseq}) and ({\ref{kspot})
is obtained by minimizing $E_{tot}^{OEP}[\{u_i\}]$, i.e.
\be
 \frac{\delta E_{tot}^{OEP}[\{u_i\}]}{\delta v_s({\bf r})} \biggm|_{v_s=v^{OEP}}
=0.
\label{oep}
\ee
As first pointed out by Perdew and co-workers Eq.~(\ref{oep}) is equivalent
to the Hohenberg-Kohn variational principle \cite{Perdew,Graborev}.

 The exact {\em exchange} part of $E_{xc}$ is known in terms of the single
particle orbitals, i.e.
\be
 E_x^{exact}[\{u_i\}]=-\frac{1}{4} \sum_{ij}^{occ} \int d{\bf r} d{\bf r}' \frac{u_i^*({\bf r})
u_j^*({\bf r^{'}}) u_i({\bf r'}) u_j({\bf r})}
{\vert {\bf r}-{\bf r}' \vert}.
\label{exchange}
\ee
We will not repeat the details of the OEP calculations here and refer the interested reader
to a recent review of the method \cite{Graborev}.
Also from now on we will restrict ourselves to the exchange-only case,
i.e. we neglect correlation completely and use Eq.~(\ref{exchange}) as approximation
to the exchange-correlation functional.
The resulting OEP integral equation is rather complicated to solve in practice \cite{KLI}. 
In the work of Krieger and co-workers the OEP integral equation is analyzed
and a simple approximation is made which reduces the complexity of the
original OEP equation significantly and at the same time keeps many of
the essential properties of OEP unchanged \cite{KLI}.
Krieger, Li and Iafrate gave an exact expression by transforming the OEP
integral equations into a manageable form. They got the following, still  exact,
expression for $v_x({\bf r})$ \cite{KLI}:
\beqn
v_x^{OEP}({\bf r})=v_x^S({\bf r})+\sum_i^{occ} \frac{\rho_{i}({\bf r})}{\rho({\bf r})}
(\overline{v}_{xi}^{OEP}-\overline{v}_{xi}^{HF}) && \\ \nonumber
+\frac{1}{2} \sum_i^{occ} \frac{\nabla [p_i({\bf r}) \nabla u_i({\bf r})]}
{\rho({\bf r})},
\label{exactvx}
\eeqn
where $v_x^S({\bf r})$ is the Slater potential given as the first term in Eq.(5)
when $g^{\lambda}=g^{\lambda=0}$
and $\rho_i({\bf r})=\vert u_i({\bf r})\vert^2$.
  Slater's potential can be written in terms of the first-order density matrix
$\gamma({\bf r},{\bf r}^{'})$ as follows
\be
 v_x^{Slater}({\bf r})= -\frac{1}{2} \int \frac{\vert \gamma({\bf r},{\bf r}')\vert^2}
{\rho({\bf r}) 
 {\vert {\bf r}-{\bf r}' \vert}} d{\bf r}'.
\ee
In Eq. (9) the summation runs over the orbital index for all the occupied orbitals up to the
highest occupied $m$th orbital (Fermi level).
The function $p_i$ is defined by 
\be
p_i({\bf r})=\frac{1}{u_i({\bf r})} \int d{\bf r}' [v_{xc}^{OEP}({\bf r}')-
v_i({\bf r}')] G_i({\bf r},{\bf r}') u_i({\bf r}'),
\ee
where $G_i({\bf r},{\bf r}')$ is the Green's function
\be
 G_i({\bf r},{\bf r}')= \sum_{j \neq i}
\frac{u_j^*({\bf r}) u_j({\bf r}')}{\epsilon_j-\epsilon_i}.
\ee
In practical applications the last term in Eq.(9) turned out to be quite small
in atomic systems and has small effect only at the atomic shell boundaries \cite{KLI}. 
Krieger, Li and Iafrate (KLI) \cite{KLI} have proposed a simple approximation
where this last term is neglected completely.
This might appear a rather crude approximation but it can be interpreted
as a mean-field approximation since the neglected terms averaged over the
ground-state density $\rho({\bf r})$ vanish.
This is not only true for finite systems with exponentially decaying densities
 \cite{KLI},
but also for infinite systems such as solids or polymers \cite{Graborev}.
The KLI-approximation to the exchange potential 
after some algebra reads
\be
v_x^{KLI}({\bf r})=v_x^S({\bf r})+
\sum_{i=1}^{m-1} \frac{\vert u_i({\bf r}) \vert^2}{\rho({\bf r})} \sum_{j=1}^{m-1} ({\bf A}^{-1})_{ij} (\overline{v}_{xj}^S-\overline{v}
_{xj}^{HF}),
\label{kli}
\enspace
\ee
\beqn
\sum_{j=1}^{m-1} ({\bf A}^{-1})_{ij} (\overline{v}_{xj}^S-\overline{v}
_{xj}^{HF}) =
\overline{v}_{xi}^{KLI}-\overline{v}
_{xi}^{HF}, \nonumber
\eeqn
where $m$ is the highest occupied orbital level.
\be
{\bf A}_{ji}=\delta_{ji}-{\bf M}_{ji},
\ee
\be
 {\bf M}_{ji}=\int \frac{\rho_j({\bf r}) \rho_i({\bf r})}
{\rho({\bf r})} d{\bf r}, i,j=1,...,m-1.
\ee
$\overline{v}_{xj}^S$ and $\overline{v}_{xj}^{HF}$ are given as follows:
\be
\overline{v}_{xj}^S= \int \rho_j({\bf r}) v_x^S({\bf r}) d{\bf r}.
\label{overvxs}
\ee
\be
\overline{v}_{xj}^{HF} 
 = -\frac{1}{2} \sum_i^{occ} \int d{\bf r} d{\bf r}' \frac{u_i^*({\bf r})
u_j^*({\bf r^{'}}) u_i({\bf r'}) u_j({\bf r})}
{\vert {\bf r}-{\bf r}' \vert},
\ee
\be
v_x^S({\bf r}) = - \frac{1}{2\rho({\bf r})} 
\sum_{i,j}^{occ} u_i^*({\bf r}) 
u_j({\bf r}) \int d{\bf r}' \frac{u_i({\bf r}') 
u_j^*({\bf r}')}{\vert {\bf r} - {\bf r}' \vert}
\label{slater}
\ee
\be
v_i({\bf r})=\frac{1}{u_i^*({\bf r})} \frac{\delta E_x^{OEP}[\{u_i\}]}
{\delta u_i({\bf r})},
\label{vxi}
\ee
The term corresponding to the highest occupied orbital $u_m$ has been excluded
from the sum in Eq.~(\ref{kli}), because $\overline{v}_{xm}^{KLI}=\overline{v}
_{xm}^{HF}$ \cite{KLI}.

 In this article we restrict ourself to different exchange-only methods, however,
the extension of these methods to exchange-correlation case is in principle
straightforward \cite{KLI}. 
The most trivial way of accounting for Coulomb correlation effects is the
use of a correlation energy functional either in its local or gradient
corrected form. 
However, recent calculations using a gradient- and orbital-dependent
functional in combination with exact exchange provided excellent results
for atoms only \cite{KLImol}.
The poor performance of this functional for molecules must be attributed to
the improper long-range component in the corresponding correlation hole
which is needed to cancel the long-range component in the exact exchange hole
(the combined xc hole is typically short-ranged). Most of the approximate
correlation functionals, which are derived from the homogeneous or
the inhomogeneous electron gas model, are suffering from incorrect 
long-range assymptotics and exhibit improper local behaviour \cite{Sule}.
Therefore the extension of OEP based methods to include electron correlation
will probably be the subject of further studies in the next future \cite{Sule3}.

 In this paper we do not attempt to solve Eq.~(\ref{kli}) for polymers but instead
choose a different approach proposed by Gritsenko at. al \cite{Gritsenko}.
 For the constants $w_j=\overline{v}_{xj}^{KLI}-\overline{v}_{xj}^{HF}$ in Eq.~(\ref{kli}) an alternative
expression is proposed \cite{Gritsenko} in terms of orbital energies
$\epsilon_j$. It follows from gauge invariance requirements, proper scaling and short range behaviour of the response part of Eq.(5) and Eq.~(\ref{kli}) that $w_j$ does 
only depend on energy differences. Therefore
Eq.~(\ref{kli}) turns into the following much simpler formula:
\be
 v_x^{SSP}({\bf r})=v_x^S({\bf r})+\frac{8 \sqrt{2}}{3 \pi^2} \sum_{i=1}^{m-1} 
  \frac{\vert u_i({\bf r})\vert^2}{\rho(
{\bf r})}
 \sqrt{\epsilon_F-\epsilon_i}
\label{ssp}
\ee
where $\epsilon_F$ is the Fermi level (highest occupied energy level).
We use the notation SSP (Slater's potential + step potential) for this exchange potential.
The constant 
 $\frac{8 \sqrt{2}}{3 \pi^2}$ is obtained from the homogeneous electron
gas (HEG) model so that, Eq.~(\ref{ssp}) is exact in the HEG limit \cite{Gritsenko}.
This constant is chosen as universal parameter for all the calculations.
It is one of the main advantages of this expression that one can avoid the tedious matrix
inversion of Eq.~(\ref{kli}) and also suggests the future applicability of a more general class of
$v_{xc}$ which can be designated by $v_{xc}([\{u_i\},\{\epsilon_i\}];{\bf r})$
\cite{Sule3}.
Note that the summation runs over all the occupied orbitals except the highest one, as
in Eq.~(\ref{kli}).
With the step potential-like second term $v_x^{SSP}$ provides a good 
approximation to the OEP exchange-potential (Eq.(9)) \cite{Gritsenko}
that possesses proper short-range behaviour and the characteristic atomic-shell
stepped structure.


\vspace{8mm} 

 {\bf 3. The implementation of the orbital dependent exchange potentials}\\

In this section we briefly give the summary of the theory
for a periodic one-dimensional polymer chain
based on gaussian lobe functions
\cite{Whitten}.

In the following we want to describe quasi one dimensional helical
chain polymers, i.e. we consider a single polymer chain (extending to
infinity) which may or may not have a helix structure (for general background on the LCAO method on helical polymers see, e.g. 
\cite{LDApolymer}). Mathematically this symmetry can be expressed by a screw operation
{\bf S} in terms of a translation $a$ along the $z$-axis combined with a
rotation of angle $\Theta$ about the same axis, i.e. formally,
\be
{\bf S}(a,\Theta){\bf r}=
\left( \begin{array}{c}
x cos \Theta-y sin \Theta\\
x sin \Theta+y cos \Theta\\
z+a
\end{array}
\right)
\ee
The symmetry group generated by the screw operation is an Abelian group and therefore
its irreducible representations are one-dimensional. Therefore the single-electron
wavefunctions corresponding to this problem will transform according to
\be
u_i(k,{\bf S}^n {\bf r})=exp(-ikn) u_i(k,{\bf r})
\label{transf}
\ee
where the label $k$ (which may be viewed as a normalized wavenumber) is restricted
to the range $-\pi \leq k \leq \pi$ and $i$ is simply the band index
(to avoid confusion of indexes we note that $i$ in the exponent is simply the imaginary number).

 The electrons are assumed to doubly occupy a set of one-electron orbitals,
$u_i(k,{\bf r})$, of Bloch-type. These orbitals are written as linear combinations of
$m_b$ atomic (real) basis functions. $\{\chi_{\mu},\mu=1,...,m_b\}$ 
denotes the basis atomic functions in the reference unit cell and 
$\chi_{\mu}^n({\bf r})=\chi_{\mu}^0({\bf S}^n{\bf r})$ are the corresponding
basis functions for unit cell $n$.
The ansatz for the orbitals then reads
\be
 u_i(k,{\bf r})=\sum_{n=-\infty}^{+\infty} \sum_{\mu}^{m_b}
 c_{i \mu}(k) e^{ikn} \chi_{\mu}^n({\bf r}),
\label{orbit}
\ee
which transforms according to Eq.~(\ref{transf}).
In this article the sets of index ($n,n^{'},n^{''}$), ($i,j$) and 
($\mu, \nu, \sigma, \tau$) are used, referring to cells, orbitals and contracted atomic lobe functions.
The unknown coefficients $c_{i\mu}(k)$ have to be determined through the
self-consistency procedure. Since we are dealing with an infinite chain polymer,
the sum over the neighbouring cells extends from $-\infty$ to $+\infty$.
In practice, however, we only take a finite number of neighbouring cells into
account. In section 5. we will address the question of how many cells
need to be taken into account in order to get converged ground-state energies.

 The total energy for the polymer system is expressed as 
\cite{LDApolymer}
\beqn
E_{tot}=\sum_{\mu \nu}^{m_b} \sum_{n=-\infty}^{+\infty} {\bf P}_{\mu \nu}^{0,n}
\biggm\{ -\frac{1}{2} \langle \chi_{\mu}^0 \vert \nabla^2 \vert \chi_{\nu}^n \rangle
&& \nonumber \\
+\langle \chi_{\mu}^0 \vert e_{xc}([\{u_i\}];{\bf r}) \vert \chi_{\nu}^n \rangle \biggm\}
+\frac{1}{2} \sum_{n'=-\infty}^{+\infty} \biggm{\{}
\biggm( \sum_{rs} \frac{Z_r Z_s}{\vert {\bf R}_r^0-{\bf R}_s^{n'} \vert} \biggm)
&& \nonumber \\
+\sum_{\mu \nu}^{m_b} \sum_{n=-\infty}^{+\infty} {\bf P}_{\mu \nu}^{0,n}
\biggm(\sum_{\sigma \tau}^{m_b} \sum_{n^{''}=-\infty}^{+\infty} 
{\bf P}_{\sigma \tau}^{n',n^{''}}
 \langle \chi_{\mu}^0 \chi_{\nu}^n \vert \chi_{\sigma}^{n'} \chi_{\tau}^{n^{''}} \rangle
&& \nonumber \\
-2 \sum_r \biggm{\langle} \chi_{\mu}^0 \biggm|
\frac{Z_r}{\vert {\bf r}-{\bf R}_r^{n^{''}} \vert} \biggm|
\chi_{\nu}^n \biggm{\rangle} \biggm) \biggm{\}} ,~~~~~~~~~~~~~~~
\eeqn
where $Z_r$ and ${\bf R}_r$ denote the nuclear charges and coordinates within
a single unit cell, ${\bf R}_r^n$ denotes the nuclear coordinates in
unit cell $n$ (${\bf R}_r^n=S^n {\bf R}_r$).
In our particular case the orbital dependent exchange-correlation energy density is $e_{xc}([\{u_i\}];{\bf r})=e_x([\{u_i\}];{\bf r})$ and
also $E_x=\int e_x([\{u_i\}];{\bf r}) d{\bf r}$ \cite{Sule}.

 According to Eq.~(\ref{exchange}) $E_x^{OEP}$ can be given in terms of one-particle density matrixes as well.
\beqn
E_x^{OEP}=
 -\frac{1}{4} \sum_{\mu \nu} \sum_{n,n',n^{''}=-\infty}^{+\infty}{\bf P}_{\mu \nu}^{0,n} \sum_{\sigma \tau}  
 {\bf P}_{\sigma \tau}^{n',n^{''}} ~~~~~~~~~~~~~~~~~~~~~ \\ \nonumber 
 \int \frac{\chi_{\mu}^0({\bf r}) \chi_{\nu}^n({\bf r}') \chi_{\sigma}^{n'}({\bf r}) \chi_{\tau}^{n^{''}}({\bf r}')}
 {\vert {\bf r}-{\bf r}' \vert} d{\bf r}' d{\bf r},~~~~~~~~~~~~~~~~~
\eeqn
where $\chi_{\mu}^{0}$,
 $\chi_{\nu}^n$,$\chi_{\sigma}^{n'}$ and $\chi_{\tau}^{n^{''}}$ are the basis functions in which the orbitals are expanded. The upper indexes $0$ and $n,n',n^{''}$ denote the localization of
the basis function in the reference cell and in the infinite system,
respectively.
The density matrix elements for one-dimensional periodic systems
are given by,
\be
 {\bf P}_{\sigma \tau}^{n',n^{''}}= \sum_{j=1}^{occ} \int_{-\pi}^{\pi} \frac{dk}{2\pi} c^{*}_{j \sigma}(k) c_{j \tau}(k) exp[ik(n'-n^{''})]. 
\ee
The density matrix ${\bf P}_{\mu\nu}$ is computed at each iteration
by numerical integration over the occupied part of the first Brillouin zone of 
the polymer.
 The interactions of the "reference cell" with neighbours (finite neighbouring interaction)
are taken into account by the summation over the cell index $n$. 
 The electron density can than be written as
\be
\rho({\bf r})=\sum_{n=-\infty}^{+\infty} \sum_{\mu \nu}^{m_b}
{\bf P}_{\mu \nu}^{0,n} \chi_{\mu}^0({\bf r}) \chi_{\nu}^n({\bf r}).
\ee

 The contracted gaussian-lobe $\chi_{\mu}({\bf r})$ basis functions in the reference unit cell \cite{Whitten} is equal to 
\be
 \chi_{\mu}({\bf r})=\frac{1}{\sqrt{N}} \sum_{\alpha} d_{\mu \alpha} exp(-\alpha_{\mu \alpha} \vert {\bf r}-{\bf A}_{\mu \alpha}-{\bf R}_s \vert^2),
\ee
where the summation runs over contraction index $\alpha$.
The constants $d_{\mu \alpha}$ are fixed according to the contraction pattern.
The basis functions $\{\chi_{\mu}({\bf r}), \mu=1,...,m_b\}$ are also chosen
to be normalized and
${\bf R}_s$ is the atomic position in the unit cell from where
the lobe functions are displaced.
In the basis set the higher angular momentum terms $p, d, ...$ are treated by linear
combinations of s-functions (lobes) with its origins displaced from
the nuclear position by vectors ${\bf A}_{\mu \alpha}$. 
 (for further details see Ref. \cite{Whitten}).
For instance, a $p_x$ orbital can be written as a difference between two lobe functions
$u_{p_x}=1/\sqrt{N}(u_1-u_2)$ so that the exponent in the basis set will be
\be
 \vert {\bf r}-{\bf A}_{\mu \alpha}-{\bf R_s} \vert^2=[r_x+(-1)^l A-R_{sx}]^2+\sum_{i=y,z} (r_i-R_{si})^2.
\ee
where
$A$ is the magnitude of the displacement of the lobe centers, and $R_{si}$ is the component of the position vector of 
nucleus $s$.
 According to Eq. (18), the Slater potential 
in terms of gaussian basis sets and of the
self-consistently determined density matrix can be written 
\beqn
v_x^{Slater}({\bf r})= -\frac{1}{2 \rho({\bf r})} \sum_{n,n',n^{''}=-\infty}^{+\infty} \sum_{\mu \nu} {\bf P}_{\mu \nu}^{0,n}~~~~~~~~~~~~~ && \\ \nonumber \sum_{\sigma \tau} 
 {\bf P}_{\sigma \tau}^{n,'n^{''}} 
 \chi_{\mu}^0({\bf r}) \chi_{\sigma}^{n'}({\bf r})   
 \int \frac{\chi_{\nu}^n({\bf r}^{'}) \chi_{\tau}^{n^{''}}({\bf r}')}
 {\vert {\bf r}-{\bf r}' \vert} d{\bf r}'. ~~~~~~~~~
\eeqn
While the integration with respect to the primed ${\bf r}^{'}$ coordinate can be carried
out analytically \cite{Shavitt},
the second integration to calculate the matrix elements according to Eq.~(\ref{overvxs}) can be accomplished 
only numerically which makes the procedure somewhat time consuming.
However, further attention will be devoted to simplifying this
integral using auxiliary fit functions \cite{LDApolymer}.\\

Since we are dealing with a Hartree-Fock-Roothan-Hall SCF-LCAO-MO approximation  in a basis set representation \cite{Szabo}, one has to apply
the Fock matrix formalism and has to form the corresponding Fock matrix
elements:
\be
 {\bf F}_{\mu \nu}^n={\bf H}_{\mu \nu}^n+{\bf J}_{\mu \nu}^n+{\bf F}_{\mu \nu}^
{n,x},
\ee
where
 ${\bf H}_{\mu \nu}^n,{\bf J}_{\mu \nu}^n$ and ${\bf F}_{\mu \nu}^
{n,x}$ are the one-electronic, Coulomb and exchange contributions to the
Fock matrix ${\bf F}_{\mu \nu}^n$ with $n$ unit cell so that
\be
{\bf H}_{\mu \nu}^n=\langle \chi_{\mu}^0 \vert -\frac{1}{2} \nabla+v_{ext}({\bf r}) \vert 
\chi_{\nu}^n \rangle,
\ee
\be
{\bf J}_{\mu \nu}^n=\langle \chi_{\mu}^0 \vert v_H({\bf r}) \vert \chi_{\nu}^n
 \rangle,
\ee
and
\be
{\bf F}_{\mu \nu}^{n,x}
=\langle \chi_{\mu}^0 \vert
v_x({\bf r}) \vert \chi_{\nu}^n \rangle.
\ee
According to Eq.~({\ref{kseq}) and Eq.~(\ref{orbit})
the stationary one-electron periodic wavefunctions $\{u_i(k,{\bf r})\}$ are obtained
in the usual way by solving the eigenvalue equations
\beqn
\sum_{\mu} c_{j\mu}(k) \biggm\{ \sum_{n=-\infty}^{+\infty} e^{ikn}[{\bf F}_{\mu \nu}^n 
-\epsilon_j(k) {\bf S}_{\mu \nu}^n] \biggm\}=0, && \\ \nonumber \mu,\nu=1,2,..., m_b
\eeqn
with
\be
 {\bf S}_{\mu\nu}^n= \int d{\bf r} \chi_{\mu}^0({\bf r}) \chi_{\nu}^n({\bf r}),
\ee
at a finite number of $k$ points in the Brillouin zone (BZ). This gives the
band structure ${\epsilon_j(k)}$ and the eigenvectors.

\vspace{6mm}

{\bf 4. Computational details}

\vspace{6mm}

 On the basis of the formalism described above the total energy  
and the electronic band structure have been calculated for polyethylene
by various exchange-only methods. All of these schemes can be considered
as orbital dependent methods and results are compared with other
density functional exchange-only methods like the simple exchange-only LDA ($X\alpha$) procedure.
Hartree-Fock calculations have been carried out as well to make
valuable comparison with the previous methods.
For all the calculations a new code is used which has been developed in our 
group and the results are compared with other polymer codes like
Mintmire's program \cite{LDApolymer}.
This new code is based on the Erlangen periodic Hartree-Fock program \cite{Erlangen} 
and has been developed by S. Kurth and P. S\"ule \cite{diogenes}.
During the calculations the number of $k$-points and the convergence criteria are set to $25$ and
$10^{-5}$ (the change of the density matrix elements from one iteration to the next), respectively.
  For polyethylene the experimental "zig-zag" geometry is used \cite{LDApolymer}.
Since the various properties of polyethylene depend significantly on the geometry, we give
the structure we have used for the calculations: the C-C and C-H bond lengths 
and
the CCC, HCH, HCC bond angles are set to 
$1.54 \AA$, $1.10 \AA$ and $113$, $108$ and $113$ degrees, respectively.
Recent geometry optimization calculations with various DFT functionals provided values
in very good agreement with these experimental bonding parameters \cite{Hirata}.
Three types of basis sets are used: Clementi's minimal (7S/3P), the larger double-$\zeta$ (9S/5P) and (11S/7P) basis sets
 for carbon and the 4S, 6S and 8S for the hydrogen atom, respectively \cite{Clementi}.
  To make valuable comparison with previously published atomic and molecular results \cite{KLIatom,KLImol} and also to check
the reliability of our code we have made calculations 
with a translation vector of about $20.0$ a.u. and with
the minimal number of neighbouring interactions ($neig=1$). Using these parameters the computed
properties must be very close to those obtained by atomic or molecular codes.
The test provided nice agreement with atomic calculations obtained by Hartree-Fock,
Slater's or SSP exchange-only method which confirms the reliability of our code.

\vspace{6mm}

{\bf 5. Results and discussion}\\

\vspace{6mm}

 The comparison is carried out for Be and for LiH \cite{KLImol,Gritsenko} (Table I)
in the above-mentioned atomic and molecular limit with our polymer code
\cite{diogenes}.
 The deviation
of the total energies from HF can be compared with results obtained by others \cite{Gritsenko}:
we found 12, 5 mHartree for Slater and SSP, while Gritsenko {\em et al.} obtained 11 and 0.0, 
respectively.
For LiH the deviations: 
0.0 , 0.0 mH, while Grabo {\em et al.} \cite{KLImol} found 6 mH for Slater.
The discrepancy between our and other's results can be considered rather small and is
probably
due to the different basis set applied here. 
The total energies for $C_2H_4$ are compared as well in the molecular limit
and the following values are obtained with Hartree-Fock, Slater and SSP exchange-only
methods: -77.61066, -77.58735 and -77.60790 a.u.
One can see that the deviation from HF is 24 mH for Slater and only 3 mH for SSP.
For the highest occupied orbital energies we got -.4273, -.4878 and -.3854 a.u.
by HF, Slater and SSP.
For LiH the corresponding numbers are as follows: -.2914, -.3228 and -.3088 a.u.
which are compared with Grabo's values \cite{KLImol}, -.3017 (HF), -.3150 (Slater) and -.3011 (KLI).
We also compared our calculated HF total energy and Fermi level with those of obtained by other
{\em ab initio} codes like the Gaussian 94 package \cite{G94} for LiH using
the STO-3G minimal basis set. The agreement is acceptable which also indicates
that our code works in a proper way.

 Our results for some physical properties of polyethylene are shown in Table II 
and III.
The results are listed in terms of the number of the neighbouring interactions.
Actually, it turns out that at least 5 neighbours are necessary to get converged
results for all the physical properties we are interested in. For comparison
we have also calculated the properties for $8$ neighbouring cells.
In Table III values are given obtained from Mintmire's polymer code \cite{LDApolymer} as well,
which is using the multiple expansion technique for the neighbouring
interaction and can therefore be considered as an infinite order approach \cite{LDApolymer}.
However, in this code a different basis set is used (721/51/1*) which makes the comparison somewhat more difficult.

 The comparison of total energies shows that Hartree-Fock provides the lowest 
energies while the various Kohn-Sham schemes result in somewhat
higher energies. 
OEP leads to total energy values that are upper bounds to the HF results and
generally the inequality, $E_{HF} < E_{OEP} < E_{KLI} < E_{LSDX}$ holds
\cite{KLI}. 
Actually the difference between HF and SSP is 10 mH. For small diatomic molecules T. Grabo and
E. K. U. Gross \cite{KLImol} found a difference of about 10 mH between KLI and HF as well, which certainly
indicates that the SSP potential given by Eq.~(\ref{ssp}) is not a bad alternative for KLI. 
For a larger basis set the SSP and HF total energies differ more significantly ($14$ mH), which
is however, much smaller than the corresponding difference for Slater ($44$ mH).
We believe that the bulk part of the difference is not due to the approximation
of the exchange potentials applied here, but to the different nature of Hartree-Fock
and DFT approaches in accordance with earlier studies on molecules \cite{KLI,KLImol}.
 However, in a polymer chain calculation we get for the
one electronic energies quite significant deviation from the Hartree-Fock values (Table II-III). 
There is a trend in these results: the employment of exchange potentials results
in the raising of the Fermi level and the falling of the first virtual levels
giving a smaller gap when compared with Hartree-Fock for polyethylene.
$\epsilon_m^{HF}$ remains almost constant with respect to the increase of
the number of neighbours taken into account in the
neighbouring approach.

 Neglect of the response part in Eq.~(\ref{kli})  leads to somewhat higher energies for $E_{tot}$.
Thus the addition of a repulsive response part to the attractive bare Slater potential brings
$E_{tot}$ and $E_x$ closer to the corresponding Hartree-Fock values \cite{Gritsenko}.
The error in $E_{tot}$ obtained with the naked Slater potential $v_x^S$ increases
with the basis set and reaches $20 mH$ for the largest basis set when compared
with SSP.

   Table II-III represent exchange $E_x$ energies as well. Slater yields
too negative $E_x$ because of the attractive character of $v_x^S$.
The addition of the approximate repulsive step potential (SSP) brings the
$E_x$ values much closer to $E_x^{HF}$ but still remains somewhat more negative (Table III). 
Gritsenko {\em et al.} \cite{Gritsenko}  found, however, the overcompensation of $E_x^{SSP}$ values 
compared with $E_x^{OEP}$ for atoms. 
The inequality $E_x^{OEP} \leq E_x^{HF}$ \cite{LP} is clearly reproduced
by Slater and SSP with the larger basis set. 

 The most striking feature to be mentioned is that the  
highest occupied orbital energies $\epsilon_m$ obtained by SSP differ from HF values significantly,
and are less negative than $\epsilon_m^{X\alpha}$.
The values $\epsilon_m^{Slater}$ are within the range of $[\epsilon_m^{HF};\epsilon_m^{X \alpha}]$.
These results are rather surprising since atomic and molecular calculations show that
the $\epsilon_m$ values obtained by HF or OEP are close to each other \cite{KLI,KLIatom,KLImol}.
While this comes as a surprise we consider the raising of the Fermi level due to the
periodic effects, which come into play only, however, when one employs localized exchange-potentials for
infinite systems.
One can see how the periodic effect comes into play when the change of $\epsilon_m$ 
is examined in terms of the order of the neighbouring interactions in Table II.
As may be read from Table II, when $neig=1$ (which case is closer to the molecular limit than to a periodic system) , $\epsilon_m$ obtained by Slater or SSP
is rather close to the HF value, however, when $neig$ is increased the Fermi level 
shifted upward and is
getting closer to zero progressively. This phenomenon is clearly the manifestation of
periodic effect on highest occupied orbitals. Hartree-Fock does not produce such a
phenomenon since it exhibits non-local orbital dependent exchange-potential, which is always deeper than the local counterpart,  and thus
is keeping the Fermi level at a deeper energy level. 
Table II also reports that with one neighbour ($neig=1$) the KS-based
methods exhibit too wide gap and as the periodicity is building up progressively, the gap is lowered significantly.

 The $X\alpha$ results differ more significantly from all the other
methods, yielding higher total energies and higher Fermi levels which is due to the
wrong exponential long range behaviour of the LDA exchange potential.
The smallest HOMO-LUMO transition, the fundamental (eigenvalue) band gap is found with all
the methods at the edge of the Brillouin-zone ($k=1$).
As can be seen in Table III,
increasing the basis set leads to a decrease of the calculated gap for all the methods, which is
mainly due to the low first virtual levels when compared with HF.
While the HF gap is still too high the calculated Slater's, SSP  gaps are too low.
$X\alpha$ provides value surprisingly close to the experimental one
\cite{LDApolymer} while in the literature $X\alpha$ is known for typically giving
gaps which are too small \cite{Bylander}. 
It is worth to note, however, that the HF gap goes through significant changes
with the size of basis set as well in the $[14,24]$ eV range \cite{Ladik}.
It is mainly due the downwards movement of
the HF first virtual level $\epsilon_{m+1}$ while 
the $\epsilon_m$ level remains almost unchanged with
respect to the size of the basis set.
As is well-known, Hartree-Fock provides too large gaps \cite{Ladik}.
Although Slater and SSP give values around
the experimental one ($\sim 8.8 eV$ \cite{LDApolymer,expgap}) with the minimal basis set (Table II)
the gap obtained by SSP is closer to the $X\alpha$ value.
However, the band gap is narrowed significantly when the basis set is enlarged (Table III).
 This clearly must be attributed to the downwards movement of the first
virtual levels $\epsilon_{m+1}$ and at the same time to the raising of the Fermi level
$\epsilon_m$. 

 To understand the "small eigenvalue gap" problem
 in Fig. 1 and 2 the band structure is plotted obtained by various methods using
the Clementi's minimal and double-$\zeta$ bases (Fig. 1 and 2, respectively).
Analyzing and comparing systematicaly the band structures in Fig 1-2 one can see that "new" virtual bands
appear below the first virtual level obtained with the minimal basis set.  
The larger the basis set is, the larger the virtual space becomes which can
result in the appearance of virtual states with low orbital energies.
We would like to emphasize that the appearance of new low-lying virtual levels below the
ones of the minimal-basis calculations is certainly not unexpected. 
As the basis set increases new
states appear among the virtual canonical orbitals with orbital energies that form a dense subset of possible
virtual energies
\cite{Bartha}. 
It is not easy then to separate out the virtual state which corresponds
to the lowest conduction band state
and also the one-electron excited states are hidden in the virtual subspace.

 A more serious aspect of the small gap problem to be considered is that one has to take into account
the derivative discontinuity to get the right gap when using OEP \cite{DFT,Graborev}.
For continuum approximations to $E_{xc}$ like LDA or GGA this discontinuity
vanishes. In OEP and in our approaches, however, we have a finite derivative
discontinuity \cite{KLI,Graborev}. 
In OEP the exact band gap can be written as follows \cite{Graborev,Stadele},
\beqn
 \Delta=\Delta_{nonint}^{KS}+\Delta_{xc}~~~~~~~~~~~~~~~~~~~~ \\ \nonumber 
 = \epsilon_{\nu}^{KS}(N)-\epsilon_i^{KS}(N)+\Delta_{xc}~~~~,
\eeqn
where $N$ is the number of electrons. One has to add the xc-contribution 
($\Delta$ is identical with the band gap) to the difference of the
Fermi ($i$th level) and first virtual KS one-electron energies (level $\nu$) which represent 
the highest valence band and the energetically lowest conduction band states, respectively.
$\Delta_{nonint}^{KS}$ is the Kohn-Sham eigenvalue band gap.
The exchange-only contribution to the exact band gap $\Delta_x$ can be given \cite{Stadele}
\beqn
\Delta_x(i \rightarrow \nu)=\langle u_{\nu} \vert \hat{v}_x^{HF}-\hat{v}_x^{OEP}[\rho] \vert u_{\nu} \rangle 
 \\ \nonumber 
 - \langle u_i \vert \hat{v}_x^{HF}-\hat{v}_x^{OEP}[\rho] \vert u_i \rangle
\\ \nonumber
-\langle \nu i \vert \nu i \rangle+2 \langle \nu i \vert i \nu \rangle,~~~~~~~
\eeqn
with $\hat{v}_x^{HF}$ being the nonlocal HF exchange operator constructed
from the $N/2$ occupied KS orbitals.
This formula is coming from the first order DFT perturbation theory \cite{Gorling}
and the eigenvalue gap $\Delta_{nonint}^{KS}$ represents the zeroth order term in
the perturbation expansion \cite{Stadele}.
 We calculated $\Delta_x$ using Slater and SSP methods and the results can be
seen in Table II-III. Values are also obtained when the numbers of the neighbours
are minimal ($neig=1$). In Eq. (38) the last terms will vanish when $neig \rightarrow \infty$.
Therefore terms
 $\langle \nu i \vert \nu i \rangle$ and $\langle \nu i \vert i \nu \rangle$
will vanish for systems with periodic boundary conditions in the limit of an
infinite number of unit cells \cite{Stadele}.
We have studied the convergence of $\Delta_x$ and as far as the minimal basis set is concerned we found no
significant change in the magnitude at $neig=5$.
The discontinuity is guaranteed to be smaller than the true band gap by
its definition. This is reproduced by our calculations only with small basis set.
Calculations with the small basis set resulted in negative value for
$\Delta_x$ which corrects the too large KS band gaps in the right way (Table II.).
With the larger basis sets $\Delta_x$ becomes positive and roughly
represents the twice of the KS band gap.
This is in accordance with the finding 
of Stad\"ele {\em et al.} \cite{Stadele}
that
the corrected band gaps are close to the Hartree-Fock gap.
 SSP gives the value of $12.7$ eV for the true band gap which is really close
the HF gap ($15$ eV). 
As Stad\"ele {\em et al.} has pointed out, $\Delta_x$ equals the difference
between the HF and the OEP eigenvalue gaps, if we assume that the difference
between the HF and OEP orbitals is negligible \cite{Stadele}.
Therefore, it is not surprising, that such an approximation to exchange-only OEP, like the SSP
approach, provides band gap very similar to the HF one.

 We got the corrected gap $6.9$ eV using the Slater's approach which is rather close to the
X$\alpha$ value $6.8$ eV when the basis
(9S/5P) is used.
Further increase of the basis set (11S/9P) results in even larger gap $\Delta$
for all the methods considered in this article.
X$\alpha$ LDA provides a gap $7.6$ eV close to the experimental value \cite{LDApolymer}.
This is in accordance with the expectations since a continuum functional,
like the $X\alpha$ functional, which averages over the discontinuity, should
yield a gap that is reasonably close to the experimental one.
The discrepancy between the calculated gaps and the experimental one must be partly attributed to the
correlation contribution to the derivative discontinuity ($\Delta_c$).
Although we would not like to go into speculations, however, it is quite
probable that $\Delta_c$ will provide much smaller contribution to
the true band gap than $\Delta_x$. Further attention must be given
 to this in order to determine the magnitude of $\Delta_{xc}$ precisely.
Nevertheless, the naked Slater exchange potential provides a corrected band gap $\Delta=9.7$
eV in much better agreement with the experiment ($\simeq 8.8$ eV) when
the large (11S/7P) basis set is used (Table III.).

 The effect of electron correlation on the gaps also should be carefully 
examined, however the recent publications indicate only minor importance of correlation
in DFT gap calculations \cite{LDApolymer}. This may well be due to 
the small effect of LDA correlation potential on the eigenvalues \cite{Bylander}. 
Because the correlation energy is strongly dependent not only on occupied states but on the virtual states as well, no simple density functional can be
expected to yield coherent gaps for different systems.
Only those correlation energy functionals 
can provide reasonable eigenvalues and can be checked against experimental
data by any chance,
which are treated strictly at the orbital dependent level of theory.

 The comparison of the Slater and SSP valence energy bands with those of X$\alpha$
and HF again shows that the shape of the bands does not change very much
 (Fig 1-2).
But the two highest valence energy bands are much further apart than in
X$\alpha$ or HF.
This is at least in part responsible for the small gaps we get for Slater and SSP.
 As already mentioned above, 
 comparing Fig. 1 with Fig. 2 one can see that using larger basis set (Fig 2.) virtual
bands are appearing below the zero energy level. This phenomenon
is clearly demonstrated by all the Kohn-Sham based methods except $X\alpha$.
Note that the lower three curves in both figures are the occupied valence energy levels 
(the deepest, core level is not plotted). 
The virtuals of HF do not go below zero as well. However, the first virtual levels are 
lowered significantly as well so that the band gap is reduced from 22 eV to
15 eV. 
The highest three virtual levels in Fig. 2 are similar to the virtuals
obtained by the minimal basis set (Fig. 1).
The smallest energy difference from the highest occupied level to the first virtual is 
at the edge of the Brillouin zone ($k=1$), however, around $k=0.5$ one can 
see quasi-degeneracy of certain bands or even crossing of virtual levels
close to the edge of the Brillouin zone (Fig. 2.).
 Comparing Fig 1 and Fig. 2 it is obvious that increasing the basis
set shifts the valence bands upwards but does not change their shape and
relative position very much.
But from the same figures one can see that the shape and position of the 
virtual bands changed a lot when increasing the basis set.
The energy bands get shifted upwards as one moves from Slater to SSP.
This seems to be consistent with Gritsenko's and Krieger's work:
in Gritsenko's Fig. 4 \cite{Gritsenko} the Slater exchange potential for Ne and Mg
gets shifted upwards when the response part is added.
Krieger's Fig.2 \cite{KLI} shows the same behaviour for the Ne atom as one goes from the Slater exchange potential
 to better exchange potentials as KLI and exact OEP.
Comparison of the Slater and SSP valence energy bands with those from X$\alpha$
and HF again show that the shape of the bands does not change very much.

  The maximum of the valence band energy is appearing in the range of [-6,-11] eV with
the lowest and highest values for
$X\alpha$ and Hartree-Fock, respectively, compared to experimentally suggested
values for the ionization potential of $9.6-9.8$ eV \cite{Exp}.
However, it must be emphasized that others give lower experimental values in
the $7.6-8.8$ eV range for this property \cite{Ladik,Springbord}.
SSP provides the much lower $-4.1$ eV value (Slater: $-7.6$ eV).
By density functional linear muffin-tin orbital method the value of $-5.1$ eV is obtained
for helical polyethylene \cite{Springbord}.
The calculated width of the lowest valence band for $X\alpha$, HF, SSP and Slater's approach is, respectively
$6.2, 9.0$, $4.4$  and $3.6$ eV, which is to be compared with the
experimental value of $7.2$ eV \cite{Exp}.
The total valence bandwidth is $14.0$ eV ($X\alpha$), $15.3$ eV (Slater and SSP) and
$19.8$ eV (HF), compared to an experimental value of $16.2$ eV \cite{Exp}.
Slater and SSP perform quite well for this band width. 
The calculated gap between the lowest valence band and the minimum of the
wider valence bands is $3.3$ eV (HF), $1.9$ eV ($X\alpha$), $3.0$ (SSP) and $3.2$ eV (Slater), respectively.
These numbers have to be
compared with the experimental 2.0 eV \cite{Exp}. 
The HF and $X\alpha$ values are in accordance with those obtained by others
\cite{Ladik,LDApolymer}. 
The bottom of the valence band with $\sigma$ symmetry is roughly in the range of [-31,-22] eV, which is in qualitative
agreement with photoemission data \cite{Exp} with 2 eV difference in avarage.


\vspace{8mm}

 {\bf 6. Conclusions}\\

 Calculations have been carried out for polyethylene with a new polymer code using various orbital-dependent
exchange-only potentials.
The Slater potential as well as its improved version, which incorporates a
step-potential for the response part of the exact exchange potential, are used and the
results are compared with those obtained by Hartree-Fock or the $X\alpha$ methods.
The addition of the step-potential as the response part of the exact exchange to the Slater
potential results in deeper total energy and somewhat wider gap
for polyethylene.

  To test the quality and reliability of our code, calculations have been performed
in the atomic and molecular limit. The results agree with those
obtained by others.

 In general, we find that the band structure calculated by different $v_x$ potentials are 
similar with little qualitative difference, while all of them differ significantly from
HF which has valence bands significantly lower, especially in the deeper valence region and also exhibits higher
virtual levels.
While the band gap obtained by Clementi's minimal basis set, is close to the experiment,
 a larger basis leads to gaps which are rather small. 
However, this poor performance for the gap is attached not to the approximate
nature of the methods applied in this paper or not even to OEP but rather to
the difficulties of gap calculations in extended systems.

 Less negative Fermi levels are calculated by Kohn-Sham based orbital dependent methods than by Hartree-Fock,
which comes as a surprise, because in molecules the highest occupied energy levels are
very close to those of HF. 
 On the basis of progressive improvement of the numbers of neighbours in the neighbouring approach
the bulk part of this difference must be attributed to periodic effects.
However, it is unclear yet whether the $\epsilon_m$ values move in the right
or wrong direction and also the question whether the periodic effect on
$\epsilon_m$ has any physical relevance remains unanswered.
Slater and SSP provide eigenvalue
band gaps which are too low. In the discussion section we are speculating on the 
possible reasons for this unexpected phenomenon.
The exchange component to the derivative discontinuity is calculated in order
to correct the Kohn-Sham band gap. These calculations indicate that
 the discontinuity $\Delta_x$ is roughly twice of the Kohn-Sham eigenvalue 
band gap. 
The corrected band gaps are close to those of obtained by LDA.
On the basis of results reported here the approximate exchange-only potential SSP seems to provide a true band gap
for polyethylene which is rather close to the Hartree-Fock gap. The simple Slater exchange potential provides band gap in much better
agreement with the experiment.


\baselineskip 5mm 

\vspace{8mm}

{\scriptsize
{\bf Acknowledgments}}
{\scriptsize
The authors wish to thank Dr. Ferenc Bartha for helpful discussions. 
This work was supported by the Flemish Science Foundation.
One of us (S.K.) gratefully acknowledges financial support from
the Deutsche Forschungsgemeinschaft.\\

\vspace{1cm}


\normalsize


\begin{thebibliography}{999}


\bibitem{DFT}
W. Kohn, P. Vashishta, {\em General Density Fucntional Theory},
in.: {\em Theory of the Inhomogeneous Electron Gas}, (Plenum Press,
New York, 1983),
R. M. Dreizler, E. K. U. Gross, {\em Density Functional Theory}
(Springer, Berlin, 1990).,
\'A. Nagy, {\em Density Functional Theory and Application to Atoms and
Molecules}, Physics Reports, {\bf 298}, 1. (1998)

\bib{Sharp}
R. T. Sharp, G. K. Horton, Phys. Rev. {\bf 90}, 317. (1953)\\

\bib{TS}
J. D. Talman, W. F. Shadwick, Phys. Rev. {\bf A14}, 36. (1976)

\bib{HK}
P. Hohenberg, W. Kohn, Phys. Rev. {\bf B136}, 864. (1964)\\ 

\bibitem{KLI}
J. B. Krieger, Y. Li, G. J. Iafrate, Phys. Rev. {\bf A45}, 101. (1992);
see also in: {\em Density Functional Theory}, p. 191, Ed. by E.K.U. Gross
and R. M. Dreizler, NATO ASI Series Vol. B337, 1995\\ 

\bibitem{KLIatom}
E. Engel, S. H. Vosko, Phys. Rev. {\bf A38}, 3098. (1993);
Y. Li, J. B. Krieger, G. J. Iafrate, Chem. Phys. Lett., {\bf 191}, 38.
(1992);
T. Grabo, E. K. U. Gross, Chem. Phys. Lett., {\bf 240}, 141. (1995)

\bibitem{KLImol}
T. Grabo, E. K. U. Gross, Int. J. Quant. Chem., {\bf 64}, 95. (1997)\\

\bibitem{Kotani}
T. Kotani, Phys. Rev. Lett., {\bf 74}, 2989. (1995)\\

\bib{Bylander}
D. M. Bylander, L. Kleinman, Phys. Rev. {\bf B52}, 14566. (1995);
Phys. Rev. Lett. {\bf 74}, 3660. (1995);
{\bf 75}, 4334. (1995)\\

\bib{Godby}
R. W. Godby, M. Schl\"uter, L. J. Sham, Phys. Rev. {\bf B36},
6497. (1987)

\bib{PW91}
J. P. Perdew, in: {\em Electronic Structure of Solids '91}, Akademie Verlag,
Berlin 1991, J. P. Perdew, K. Burke, M. Ernzerhof, Phys. Rev. Lett., {\bf 77},
3865. (1996); {\em ibid}, {\bf 78}, 1396. (1997)

\bib{Ortiz}
G. Ortiz, Phys. Rev. {\bf B45}, 11328. (1992)

\bibitem{Ladik}
J. J. Ladik, {\em Quantum Theory of Polymers as Solids},
Plenum Press, New York (1988)\\

\bibitem{LDApolymer}
J. W. Mintmire, in {\em Density Functional Methods in Chemistry};
J. K. Labanowski and J. W. Andzelm, Eds. (Springer-Verlag, New York,
1991), p. 125.;
M. S. Miao, P. E. Van Camp, V. E. Van Doren, J. J. Ladik;
J. W. Mintmire, Int J. Quant. Chem., {\bf 64}, 243. (1997);
Phys. Rev. {\bf B54}, 10430. (1996)

\bibitem{Sule}
P. S\"ule, O. Gritsenko, \'A. Nagy, E. J. Baerends, J. Chem. Phys.;
{\bf 103}, 10385., (1995), P. S\"ule, PhD thesis, {\em Electron Correlation in
Density Functional Theory: Concept and Application}, Debrecen 1996\\

\bib{Szabo}
A. Szab\'o, N. S. Ostlund, {\em Modern Quantum Chemistry} (McGraw-Hill;
New York, 1989), S. Wilson, {\em Electron Correlation in Molecules},
(Clarendron Press, Oxford, 1984)\\

\bibitem{Slater}
 J. C. Slater, Phys. Rev. {\bf 81}, 385. {1951}

\bibitem{Gritsenko}
O. V. Gritsenko, R. van Leeuwen, E. van Lenthe, E. J. Baerends,
Phys. Rev. {\bf A51}, 1944., (1995)\\

\bibitem{Leeuwen}
R. van Leeuwen, O. V. Gritsenko, E. J. Barerends,
Analysis and Modelling of Atomic and Molecular Kohn-Sham Potential,
in: {\em Topics in Current Chemistry}, Springer, 1996, Berlin\\  

\bib{Perdew}
V. Sahni, J. Gruenebaum, J. P. Perdew, Phys. Rev. {\bf B26}, 4371. (1982),
J. P. Perdew, M. R. Norman, Phys. Rev. {\bf B26}, 5445. (1982)

\bibitem{Graborev}
T. Grabo, T. Kreibich, S. Kurth, and {E.K.U. Gross},  in {\em The {S}trong
  {C}oulomb {C}orrelations and {E}lectronic {S}tructure {C}alculations:
  {B}eyond {L}ocal {D}ensity {A}pproximations}, edited by V. Anisimov (Gordon
  and Breach, Amsterdam, to appear).

\bibitem{Whitten}
J. L. Whitten, J. Chem. Phys. {\bf 39}, 349. (1966)\\


\bibitem{Shavitt}
I. Shavitt, {\em The Gaussian Function in Calculations of Statistical
Mechanics and Quantum Mechanic},
In: {\em Methods in Molecular Comput. Physics}, 1. (1963)

\bib{Springbord}
M. Springborg, M. Lev, Phys. Rev. {\bf B40}, 3333. (1989)\\

\bibitem{Erlangen}
P. Otto, {\em Integral and HFCO Program Package}, Institute for
Theoretical Chemistry, Friedrich-Alexander University, Erlangen\\

\bib{diogenes}
S. Kurth, P. S\"ule, program DIOGENES, HF, LDA and OEP polymer fortran code,
Dept. Natuurkunde, TSM, RUCA, Antwerpen\\

\bib{Hirata}
S. Hirata, S. Iwata, J. Chem. Phys., {\bf 108}, 7901. (1998)\\

\bib{Clementi}
G. C. Lie, E. Clementi, J. Chem. Phys. {\bf 60}, 1275. (1974)\\ 

\bib{Sule3}
P. S\"ule, {\em Perturbation Theory on Top of Optimized Effective Potential Method}, {\em in preparation}

\bib{Bartha}
F. Bartha, {\em private communication}\\


\bib{Exp}
K. Seki, N. Ueno, U. O. Karlsson, R. Engelhardt, E.-E. Koch, Chem. Phys.,
{\bf 105}, 247. (1986)\\

\bib{G94}
 Gaussian 94, Revision E.2,
 M. J. Frisch, G. W. Trucks, H. B. Schlegel, P. M. W. Gill,
 B. G. Johnson, M. A. Robb, J. R. Cheeseman, T. Keith,
 G. A. Petersson, J. A. Montgomery, K. Raghavachari,
 M. A. Al-Laham, V. G. Zakrzewski, J. V. Ortiz, J. B. Foresman,
 J. Cioslowski, B. B. Stefanov, A. Nanayakkara, M. Challacombe,
 C. Y. Peng, P. Y. Ayala, W. Chen, M. W. Wong, J. L. Andres,
 E. S. Replogle, R. Gomperts, R. L. Martin, D. J. Fox,
 J. S. Binkley, D. J. Defrees, J. Baker, J. P. Stewart,
 M. Head-Gordon, C. Gonzalez, and J. A. Pople,
 Gaussian, Inc., Pittsburgh PA, 1995.

\bib{LP}
M. Levy, J. P. Perdew, Phys. Rev. {\bf A32}, 2010. (1985)

K. J. Less, E. G. Wilson, J. Phys. C {\bf 6}, 3110. (1973)
\bib{expgap}

\bib{Stadele}
M. St\"adele, J. A. Majewski, P. Vogl, A. G\"orling, Phys. Rev. Lett., {\bf 79}, 2089. (1997)

\bib{Gorling}
A. G\"orling, Phys. Rev. {\bf A54}, 3912. (1996)

\bib{KLI2}
Y. Li, J. B. Krieger, G. J. Iafrate, Phys. Rev. {\bf A46}, 5453. (1992)





\end{thebibliography}


\begin{table}
\caption[]{Comparison of HF, Slater and SSP results
for Be and LiH obtained with the DIOGENES code using Clementi's minimal basis set with values from the literature.} 
\squeezetable
{\scriptsize
\begin{tabular}{lccccccc}
    &    & HF$^{our}$ & HF$^{other}$ & Slater$^{our}$ & Slater$^{other}$ & SSP$^{our}$ &
           SSP$^{other}$ \\ \hline 
Be  & $E_{tot}$/a.u. & -14.567 & -14.573 \tablenotemark[1] & -14.556 & -14.551 
\tablenotemark[2] & -14.562 & -14.560 \tablenotemark[2] \\ \hline 
LiH & $E_{tot}$/a.u. & -7.954 & -7.987 \tablenotemark[3] & -7.954 & -7.981 
\tablenotemark[3] & -7.954 & -7.987 \tablenotemark[3] \\
  & &   -7.8629 \tablenotemark[5] & -7.8620 \tablenotemark[5] & & & & \\
    & $\epsilon_m$/a.u. & -0.2914 & -0.3017 \tablenotemark[3] & -0.3228 & -0.3150  
\tablenotemark[3] & -0.3088 & -0.3011 \tablenotemark[3] \\
 & & -.2870 \tablenotemark[4] & -.2857 \tablenotemark[5] & & & & \\
\end{tabular}
}
\tablenotemark[1]\cite{KLI2}\\
\tablenotemark[2] \cite{Gritsenko}\\
\tablenotemark[3]\cite{KLImol}\\
\tablenotemark[4]{with STO-3G basis}\\
\tablenotemark[5]{with STO-3G basis using Gaussian94 \cite{G94} program package}
\end{table}

\newpage

\begin{table}
\caption[]{Calculated properties of polyethylene by various DFT methods using the 
minimal basis set\\
 {\small HF, Slater, SSP and $X\alpha$ denote the Hartree-Fock,
exchange-only method with Slater's potential (see Eq. 16), Slater's potential with orbital dependent step potential Eq.~(\ref{ssp}) and the $X\alpha$ exchange-only density functional method, respectively.
 All the properties are in a.u. except the vertical HOMO-LUMO
eigenvalue {\em gap} which is given in eV and which is computed at the edge of the Brillouin zone
 ($k=1$).
{\em Neig} is the number of the neighbours taken into account 
in the neighbouring approach. 
$E_{tot}/u$, $E_x$ are the calculated total and exchange
energies per unit cell. $\epsilon_m$ and $\epsilon_{m+1}$ are the highest
occupied and the first virtual energy levels. 
$\Delta_x$ is the exchange contribution to the total derivative discontinuity.
The calculations were carried out on the $CH_2$ unit cell.}}

\begin{tabular}{lccccc} 
 neig  &  & HF & Slater & SSP     & X$\alpha$  \\ \hline 
 1 & $E_{tot}$/u (a.u.) &  -39.03535 & -39.01888 & -39.02768 & -38.41375   \\ 
 & $E_x$ (a.u.)         & -5.82128   & -5.84774  & -5.79080  & -5.15416   \\
 & $\epsilon_m$ (a.u.)  & -.6672     &  -.4321   & -.4924    & -.5070  \\
 & $\epsilon_{m+1}$ (a.u.) & .3978   &  -.4083   & -.1084    & -.1535 \\ 
 & {\em gap} (eV)       & 29.0       & 15.1      & 16.4      & 18.0   \\  
 & $\Delta_x$ (eV)      &            & -5.7      & -4.4      & \\ \hline
 5 &                    & -38.88012  & -38.86891 & -38.87278 & -38.30351  \\
   &                    & -5.75048   & -5.76397  & -5.7379   & -5.16982  \\
   &                    & -.4747     & -.3524    & -.2312    & -.2761   \\
   &                    & .3444      & -.0433    & .1310     & .1514  \\ 
   &                    & 22.3       & 8.4       & 9.9       & 11.6  \\ 
   &                    &            & -3.5      & -1.9      &    \\ \hline
 8 &                    & -38.88008  & -38.86890 & -38.87275 & -38.30348  \\
   &                    & -5.75049   & -5.76400  & -5.7378   & -5.16983 \\
   &                    & -.4760     & -.3518    & -.2318    & -.2769 \\
   &                    & .3436      & -.0425    & .1303     & .1510 \\ 
   &                    & 22.3       &  8.4      & 9.9       & 11.6  \\ 
   &                    &            & -3.5      & -1.9       & \\ \hline
\end{tabular}
\end{table}


\begin{table}
\caption[]{Calculated properties of polyethylene by various DFT methods using the 
Clementi's double-zeta basis set (9S/5P) and with the (11S/7P) one}
\begin{tabular}{lccccc}
 neig  &  & HF & Slater & SSP &  X$\alpha$  \\ \hline 
 5 (9S/5P) & $E_{tot}$/u &  -39.01068 & -38.96683 & -38.99704 &  -38.41919   \\ 
 & $E_x$         & -5.8783    & -5.9368   & -5.8841   & -5.2527   \\
 & $\epsilon_m$  & -.4049     & -.2951    & -.1628    & -.2227  \\
 & $\epsilon_{m+1}$ & .1489   & -.1590    & .0007     & .0256  \\ 
 & {\em gap}     &  15.1      & 3.7       & 4.4       & 6.8    \\ 
 & $\Delta_x$    &            & 3.4       & 3.8       &   \\ \hline
 8 (9S/5P) &             & -39.01056  & -38.96684 & -38.99700 &  -38.4190  \\
   &             & -5.8780    & -5.9367   & -5.8832   & -5.25221  \\
   &             & -.3984     & -.2774    & -.1490    & -.2116   \\
   &             & .1558      & -.1424    & .0116     & .0375  \\ 
   &             & 15.1       & 3.7       &  4.4      & 6.8  \\  
   &             &            & 3.2       & 3.7       &      \\ \hline 
 8 (11S/7P) &    & -39.02326  & -38.99918 & -38.01359 & -38.42701  \\
            &    &   -5.8980  & -5.9486   & -5.9036  &  -5.2658   \\
            &    & -.3786     & -.2401    & -.1130   & -.2421     \\
            &    & .2320      & -.1151    &  .0755  &   .0354    \\ 
            &    &  16.6      &  3.4      &  5.1     & 7.6        \\ 
            &    &            &  6.3      &  7.6     &      \\ \hline 
 $\sim \infty$ & (Mintmire) &  & &                    & -38.45548  \\ 
 &     & &  &                                         & -5.27507    \\
 &     & &  &                                         & -.2111 \\
 &     & &  &                                         & .0750 \\
 &   & & &                                            & 7.8    \\ 
 & gap (Exp) & & &  & 8.8 \tablenotemark[1]    \\ 
\end{tabular}
{\small The notations are the same as in Table 2}\\
 \tablenotemark[1] \cite{LDApolymer,expgap}  
\end{table}


\vspace{1cm}

{\bf Fig. 1} The calculated valence band structure (eV) obtained by
various exchange-only methods as a function of the dimensionless $k$ variable with $k=0$ being the zone center and $k=1$ the zone edge. The Brillouin zone is that corresponding to the $CH_2$ unit cell. Solid lines correspond to
occupied and dashed lines to virtual levels.
The Clementi's minimal basis set is used and five neighbours are
considered for the $CH_2$ unit cell. 

\vspace{1cm}

{\bf Fig. 2} The calculated valence band structure (eV) obtained by
various exchange-only methods. Solid lines correspond to
occupied and dashed lines to virtual levels.
The two lowest dashed curves of virtual levels are of particular
interest (see text).
The Clementi's double-$\zeta$ basis set is used and 8 neighbours are
considered for the $CH_2$ unit cell.
Notations are the same as in Fig. 1

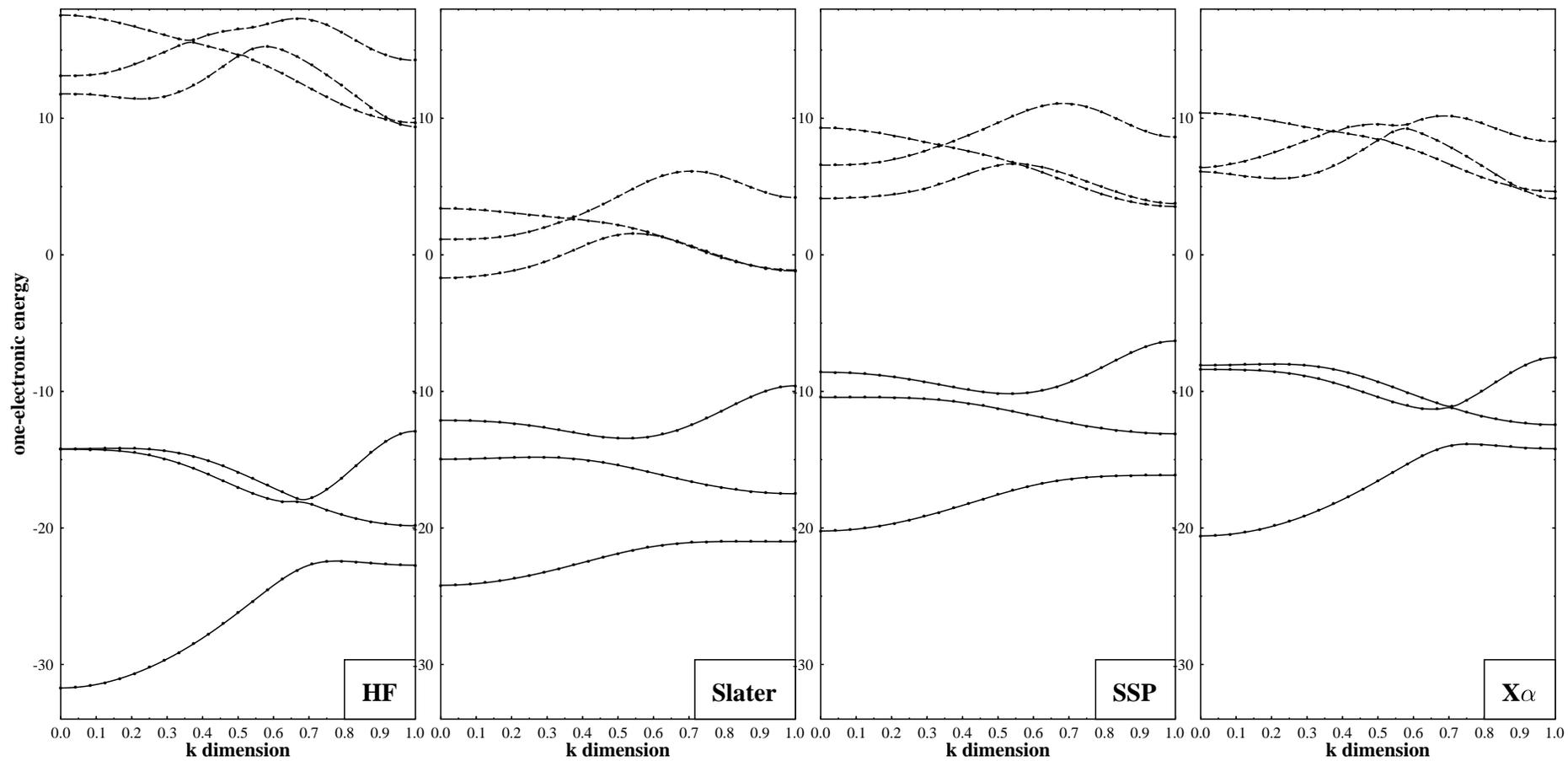
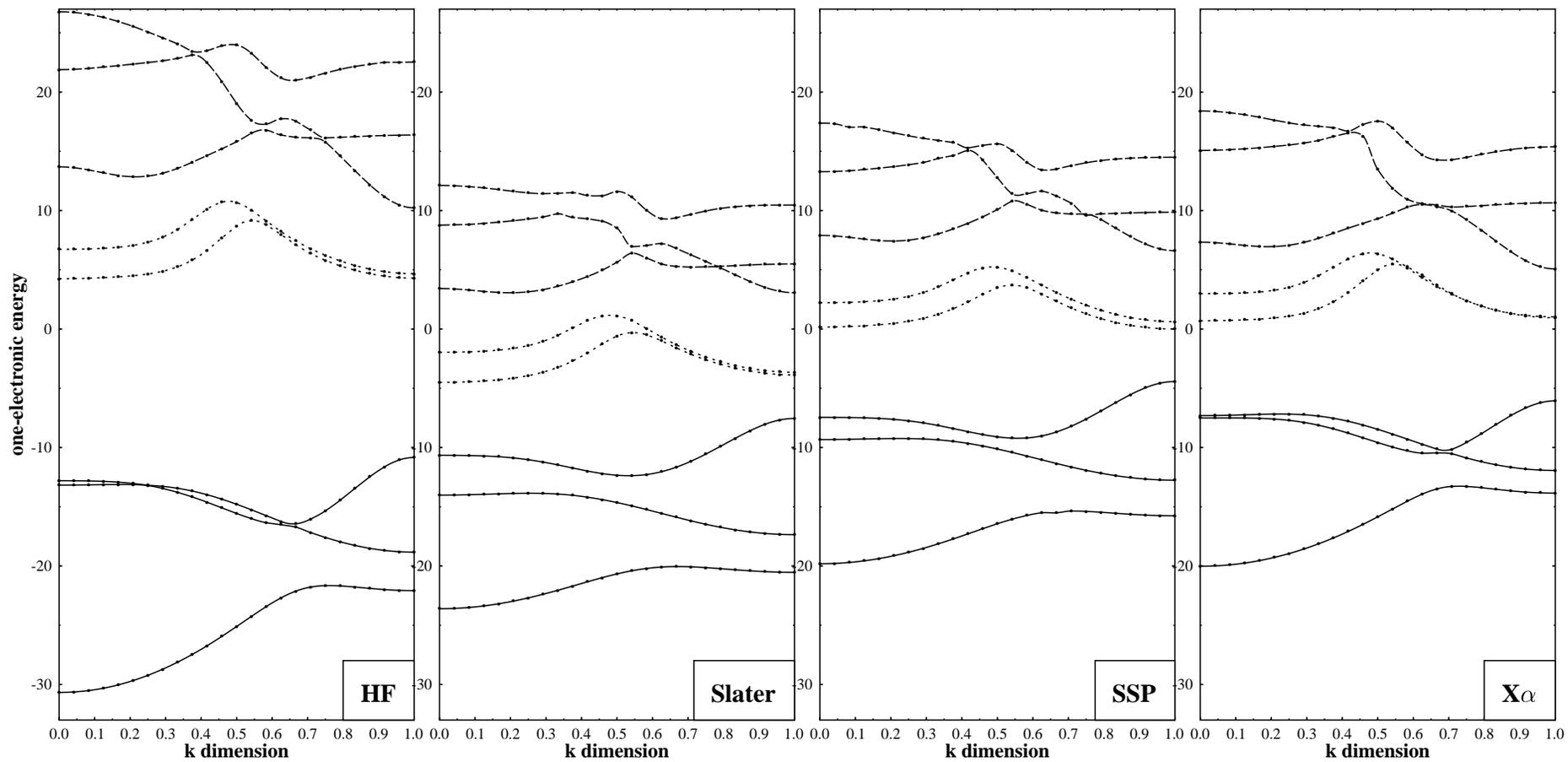
\end{document}